\documentclass[a4paper]{spie}  

\usepackage{amsmath,amsfonts,amssymb}
\usepackage{graphicx}
\graphicspath{{images/}}
\setkeys{Gin}{width=\textwidth}
\usepackage{listings}

\usepackage{upgreek}
\newcommand{\micron}{\upmu\mathrm{m}}
\usepackage{cmap} 
\usepackage[T1]{fontenc}
\usepackage[utf8]{inputenc}
\usepackage{longtable,ltcaption,array}
\usepackage[parfill]{parskip}

\usepackage{mathptmx} 
\usepackage[scaled=.90]{helvet}
\usepackage{courier}
\usepackage{alltt}
\usepackage{float}
\usepackage[english]{babel}
\usepackage{lastpage}
\usepackage{multirow}
\usepackage{tabularx}
\usepackage{supertabular}
\usepackage{fancyhdr}
\usepackage{color}
\definecolor{darkblue}{rgb}{0, 0, 0.5}
\definecolor{listingbg}{gray}{0.95}

\usepackage[colorlinks=true, allcolors=blue]{hyperref}

\title{ScopeSim: A flexible general purpose astronomical instrument data simulation framework in Python}

\author[a]{K. Leschinski}
\author[b]{H. Buddelmeijer}
\author[a]{O. Czoske}
\author[a]{M. Verdugo}
\author[b]{G. Verdoes-Kleijn}
\author[a]{W. Zeilinger}
\affil[a]{Department of Astrophysics, University of Vienna}
\affil[b]{Kapteyn Astronomical Institute, University of Groningen}

\authorinfo{Further author information: (Send correspondence to Kieran Leschinski)\\Kieran Leschinski: E-mail: kieran.leschinski@univie.ac.at}

\newcommand{\ScopeSim}{\lstinline{ScopeSim}}
\newcommand{\scopesim}{\lstinline{ScopeSim}}
\newcommand{\ScopeSimtemplates}{\lstinline{ScopeSim_templates}}
\newcommand{\scopesimtemplates}{\lstinline{scopeSim_templates}}
\newcommand{\IRDB}{\lstinline{IRDB}}

\newcommand{\Effect}{\lstinline{Effect}}
\newcommand{\Source}{\lstinline{Source}}
\newcommand{\FieldOfView}{\lstinline{FieldOfView}}
\newcommand{\Detector}{\lstinline{Detector}}
\newcommand{\DetectorArray}{\lstinline{DetectorArray}}
\newcommand{\ImagePlane}{\lstinline{ImagePlane}}

\begin{document}
\maketitle

\lstset{basicstyle=\ttfamily,
        frame=single,
        backgroundcolor=\color{listingbg},
        captionpos=b,
        showspaces=false,
        language=python}

\bibliographystyle{spiebib}

\begin{abstract}

\ScopeSim{} is a flexible multipurpose instrument data simulation framework built in Python.
It enables both raw and reduced observation data to be simulated for a wide range of telescopes and instruments quickly and efficiently on a personal computer.
The software is currently being used to generate simulated raw input data for developing the data reduction pipelines for the MICADO and METIS instruments at the ELT.
The \ScopeSim{} environment consists of three main packages which are responsible for providing on-sky target templates (\ScopeSimtemplates{}), the data to build the optical models of various telescopes and instruments (instrument reference database), and the simulation engine (\ScopeSim{}).
This strict division of responsibilities allows \ScopeSim{} to be used to simulate observation data for many different instrument and telescope configurations for both imaging and spectroscopic instruments.
\ScopeSim{} has been built to avoid redundant calculations where ever possible.
As such it is able to deliver simulated observations on time scales of seconds to minutes.
All the code and data is open source and hosted on Github.
The community is also most welcome, and indeed encouraged to contribute to code ideas, target templates, and instrument packages.

\end{abstract}


\section{Introduction}
\label{introduction}

\ScopeSim{} is a modular and flexible suite of Python packages that enable many common astronomical optical systems (observatory site/telescope/instrument) to be simulated.
The suite of packages can be used by a wide audience for a variety of purposes; from the astronomer interested in simulating reduced observational data, to a developer needing raw calibration data for testing the data reduction software.

\ScopeSim{} achieves this level of flexibility by adhering to strict interfaces between the packages, e.g: the \ScopeSim{} engine package is completely instrument and object agnostic.
All information and data relating to any specific optical configuration is kept exclusively in the instrument packages hosted in the instrument reference database (\IRDB{}).
The description of the on-sky source is kept exclusively within the target templates package (see \ScopeSimtemplates{}).
Finally, the engine makes no assumptions about what it is observing until run-time.

But why does the community need yet another instrument simulator?
Until now, most instrument consortia have developed, or are continuing to develop their own simulators\cite{hsim, schmalzl2012, simcado2016, simcado2019}.
The general consensus is that every new instrument is sufficiently different from anything that has been previously developed, that it would make little sense to adapt already existing code.
This statement is true to some extent.
Every new instrument must differ in some way from all existing instruments in order for it to be useful to the astronomical community.
However when looked at from a global perspective, every optical system is comprised primarily of elements common to many other systems.
Atmospheric emission, mirror reflectivities, filter transmission curves, point spread functions, read-out noise, detector linearity, hot pixels, are just a few of the effects and artefacts that every astronomical optical system contains.
Furthermore, while the amplitude and shape of each effect differs between optical systems, there are still commonalities in the way each effect can be described programmatically.

\ScopeSim{}'s main goal is to provide a framework for modelling (almost) any astronomical optical system by taking advantage of all these commonalities.
What \lstinline{Astropy}\cite{astropy1, astropy2} has done for the general Python landscape in astronomy, \ScopeSim{} aims to do for the instrument simulator landscape.

This paper is not intended to be a comprehensive description of the \ScopeSim{} environment.
Rather it aims to introduce the reader to the elements that make up \ScopeSim{} and directs the reader towards the online documentation for each of the packages, should the reader wish to dive deeper into the material (see Table \ref{tbl-list-of-packages}).

\section{Examples}
\label{examples}

The purpose of the ScopeSim framework is to simulate data from astronomical instruments.
The main code pattern for simulating observations with a specific instrument is the same for all use cases:

\begin{enumerate}
\item download the required instrument packages from the instrument reference database (IRDB) using \lstinline{scopesim},

\item create a description of the astronomical object using \lstinline{scopesim_templates},

\item generate a model of the desired optical system using \lstinline{scopesim} and referencing the \lstinline{irdb} package,

\item simulate and output the observed data using \lstinline{scopesim}.
\end{enumerate}

This section will illustrate how this is done with very few lines of Python code for three common science cases.
It is the authors' hope that the code snippets provided are readable by the audience of this paper.
If this is in fact not the case, the authors refer the reader to the scopesim online documentation.

\subsection{Example 1: Extended source imaging}
\label{example-1-extended-source-imaging}

\begin{figure}

\resizebox{\linewidth}{!}{\includegraphics{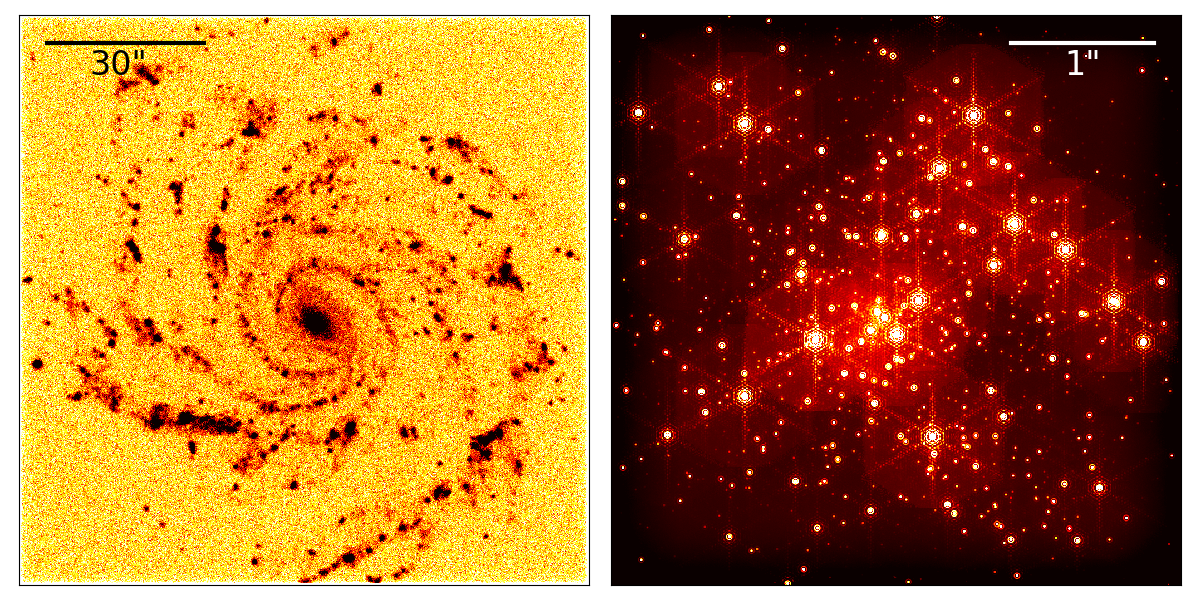}}
\caption{Left: A simulated one second observation in the Ks filter of a spiral galaxy similar to NGC\,1232L using HAWKI at the VLT
Right: A simulated one hour observation in the Ks filter of a dense $3000 M_{\odot}$ star cluster in the Large Magellanic Cloud using MICADO at the ELT.}
\label{fig:combined_1_2}

\end{figure}

The first example simulates a short (1~second) exposure with HAWKI \cite{hawki} at the VLT in No-AO mode using the Ks filter.
The target is a two component spiral galaxy using a template based on NGC\,1232L from the \lstinline{scopesim_templates} package.
The galaxy images for the old and new populations were resized to a diameter of 3~arcminutes and the associated spectra for the old and new populations \cite{brown2014} were rescaled to $12\,\mathrm{mag\,arcmin^{-2}}$ and $15\,\mathrm{mag\,arcmin^{-2}}$ respectively.
The detector window of $1024 \times 1024$ pixels covers $\sim$1.6'$\times$1.6' on sky.
The final simulated image shows primarily the star forming regions in the inner regions of the spiral arms.
The simulated detector output is shown in the left panel of Figure~\ref{fig:combined_1_2}.

This simulation setup was chosen to illustrate the noise characteristics introduced by ScopeSim.
The observation simulation requires only the following eight lines of code:

\begin{lstlisting}[frame=single]
import scopesim
from scopesim_templates.basic.galaxy import spiral_two_component

scopesim.download_packages(["locations/Paranal",
                            "telescopes/VLT",
                            "instruments/HAWKI"])

spiral = spiral_two_component(extent=180,       # arcsec
                              fluxes=(15, 12))  # mag

hawki = scopesim.OpticalTrain("HAWKI")
hawki.cmds["!OBS.dit"] = 1                      # seconds
hawki.observe(spiral)
fits_hdulists = hawki.readout()
\end{lstlisting}

The simulation work flow will be discussed in more detail in Section \ref{simulation-workflow}.
The simulation output is a FITS \lstinline{HDUList} object, containing the instrument data in the format generated by the instrument.
By default the HAWKI package produces images of size $1024\times 1024$ pixels from a fictional detector window located at the centre of the focal plane.
The package however also includes the configuration data needed to produce the standard $2\times2$ grid of $2048\times 2048$ detector images.
Switching between the two configurations is a simple matter of turning off the fictional detector window and turning on the realistic representation of the real detector array.

\subsection{Example 2: Point source imaging}
\label{example-2-point-source-imaging}

The major structures seen in the galaxy image produced in Example~1 are star forming regions.
Given the pixel-scale of HAWKI ($0.106\,\mathrm{arcsec\,pixel^{-1}}$) it would be impossible to resolve the individual stars in these regions.
MICADO\cite{micado2018} on the ELT\cite{elt}, with its 4\,mas\,pixel$^{-1}$ plate scale and adaptive optics (AO) capabilities may well be able to detect individual stars in these regions.

The following code shows how to use the ELT and MICADO (Science Team) packages to simulate observations of highly dense star cluster outside the Milky Way.
The result of this code is shown in the right panel of Figure~\ref{fig:combined_1_2}:

\begin{lstlisting}[frame=single]
import scopesim
from scopesim_templates.basic.stars import cluster

scopesim.download_packages(["locations/Armazones",
                            "telescopes/ELT",
                            "instruments/MICADO_Sci"])

cluster = cluster(mass=3e3,                     # solar masses
                  distance=50e3,                # parsec
                  core_radius=0.3)              # parsec
micado = scopesim.OpticalTrain("MICADO_Sci")
micado.cmds["!OBS.dit"] = 3600                  # seconds
micado.observe(cluster)
fits_hdulists = micado.readout()
\end{lstlisting}

This code uses the star cluster template from the \lstinline{scopesim_templates} package to create a model of a dense star cluster located in the Large Magellanic cloud ($D\sim 50\,\mathrm{kpc}$), with a core radius of 0.3\,pc and a mass of $3000\,\mathrm{M_{\odot}}$.
An exposure time of 1~hour with the Ks filter was used for the simulated observation.
This setup was chosen to show the effect of the ELT PSF on observations of densely populated fields with several bright sources.

It should be noted that the instrument package used above (\lstinline{MICADO_Sci}) is the slimmed down version of the full \lstinline{MICADO} instrument package.
Simulations using the \lstinline{MICADO_Sci} package are less computationally intensive than those which use the full \lstinline{MICADO} package (aimed at pipeline development).
The \lstinline{MICADO_Sci} package was compiled specifically for the MICADO science team to test the feasibility of various science cases with MICADO and the ELT.

\subsection{Example 3: Spectroscopy}
\label{example-3-spectroscopy}

\begin{figure}

\begin{center}
\resizebox{0.9\linewidth}{!}{\includegraphics{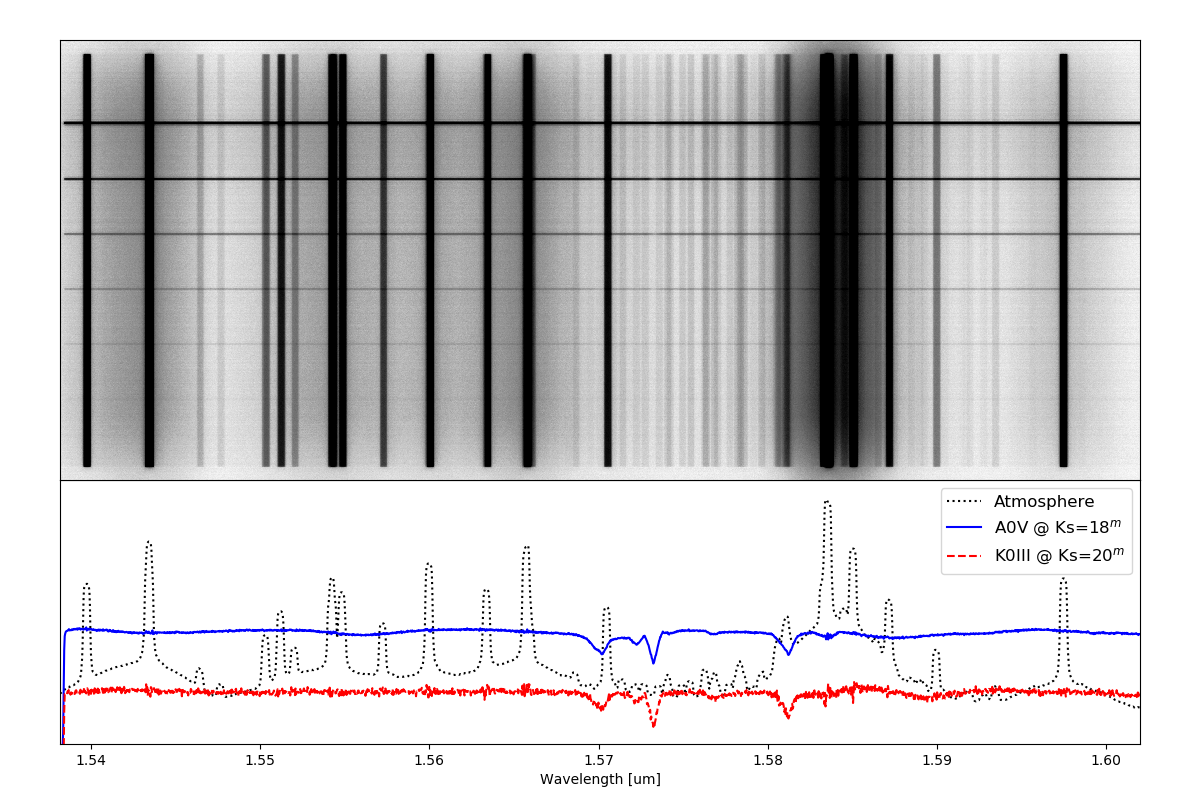}}
\caption{Top: A rectified spectral image from the MICADO detectors for a 1~hour spectrographic observation of 6~progressively fainter stars ($18\leq K_s\leq 23$).
The dark vertical bars are the atmospheric emission lines.
The thin horizontal bars are the observed stellar spectra.
The simulated wavelength range was restricted to $1.54<\lambda<1.6\,\micron$.
Bottom: Extracted spectra for the brightest ($K_s=18$), third brightest ($K_s=20$) stars, and the atmospheric background.
The atmospheric background spectrum has been subtracted from the stellar spectra.
The noise in the fainter stellar spectrum is a result of the simulated noise characteristics introduced by ScopeSim.}
\label{fig:example-3-spectra}
\end{center}

\end{figure}

The third example illustrates that ScopeSim can also be used to simulate spectroscopic observations.
While MICADO is primarily a near infrared imaging camera, it will also contain a long-slit spectrograph.
The spectroscopic mode of the \lstinline{MICADO_Sci} package allows the user to simulate reduced spectral trace data over a restricted wavelength range data -- similar to what can be expected as output from the MICADO data reduction pipeline.

The following code simulates the spectral traces of 6~stars spaced equidistantly along the long-slit aperture with magnitudes in the range $K_s=[18, 23]$.
In order to reduce computation time, the simulated wavelength range is restricted to 1024~spectral bins either side of a desired wavelength ($1.578\,\micron$).

\begin{lstlisting}[frame=single]
import numpy as np
import astropy.units as u
from scopesim import UserCommands, OpticalTrain
from scopesim_templates.basic.stars import stars

stars = stars(filter_name="Ks",
              amplitudes=np.linspace(18, 23, 6)*u.mag,
              spec_types=["A0V", "G2V", "K0III"]*2,
              x=np.linspace(-1, 1, 6),
              y=[0]*6)
cmds = UserCommands(use_instrument="MICADO_Sci",
                    set_modes=["SCAO", "SPEC"],
                    properties={"!OBS.dit": 3600,
                                "!SIM.spectral.wave_mid": 1.578,
                                "!SIM.spectral.spectral_resolution": 0.00001,
                                "!DET.height": 2048,
                                "!DET.width": 800})
micado_spec = OpticalTrain(cmds)
micado_spec.observe(stars)
micado_spec.readout(filename="basic_spectral_trace.fits")
\end{lstlisting}

As can be seen in Fig.~\ref{fig:example-3-spectra} the atmospheric emission lines are prominent in the simulated raw detector output.
The 6~stellar spectra can be seen as thin horizontal lines.
The spectra displayed in the lower panel of Fig.~\ref{fig:example-3-spectra} were extracted for the detector readout in the upper panel.
The noise in the (red) K0III spectrum is a product of the noise characteristics of the simulated observation.
These include, but are not limited to photon shot noise and electronic noise sources.

\subsection{Effects included in instrument packages}
\label{effects-included-in-instrument-packages}

The instrument packages used for these examples can be found online in the Instrument Reference Database (IRDB) Github repository (see Table~\ref{tbl-list-of-packages}).
Each package contains a description of the optical effects that are inherent to the instrument or telescope, as well as the data needed to replicate these effects.
ScopeSim allows the user to view which effects are included in the current optical model.
This example uses the \lstinline{MICADO_Sci} optical system from the previous examples:

\begin{lstlisting}[frame=single]
micado = scopesim.OpticalTrain("MICADO_Sci")
print(micado.effects)
\end{lstlisting}

During run-time ScopeSim creates an Effect object for each effect listed in the instrument configuration files.
It then applies each of these Effect objects to the on-sky Source description in turn.
Effects can be included or excluded from a simulation by using the \lstinline{.include} flag on the relevant Effect object:

\begin{lstlisting}[frame=single]
micado["readout_noise"].include = False
micado["shot_noise"].include = True
\end{lstlisting}

More information about the Effect objects is given in Section~\ref{effects-objects} as well as in the online documentation.

\section{Building blocks}
\label{building-blocks}

The ScopeSim framework has been designed to maintain strict boundaries between the simulation code, the optical model data, and the user input.
The ScopeSim framework consists of three main packages:
\begin{itemize}
\item \lstinline{ScopeSim}: the core simulation engine,

\item \lstinline{ScopeSim_templates}: a library of functions for generating descriptions of on-sky objects,

\item \lstinline{IRDB}: The instrument reference database containing the data and configuration files needed to generate the digital models of a range of telescope and instrument optical system.
\end{itemize}

\begin{figure}

\resizebox{\linewidth}{!}{\includegraphics{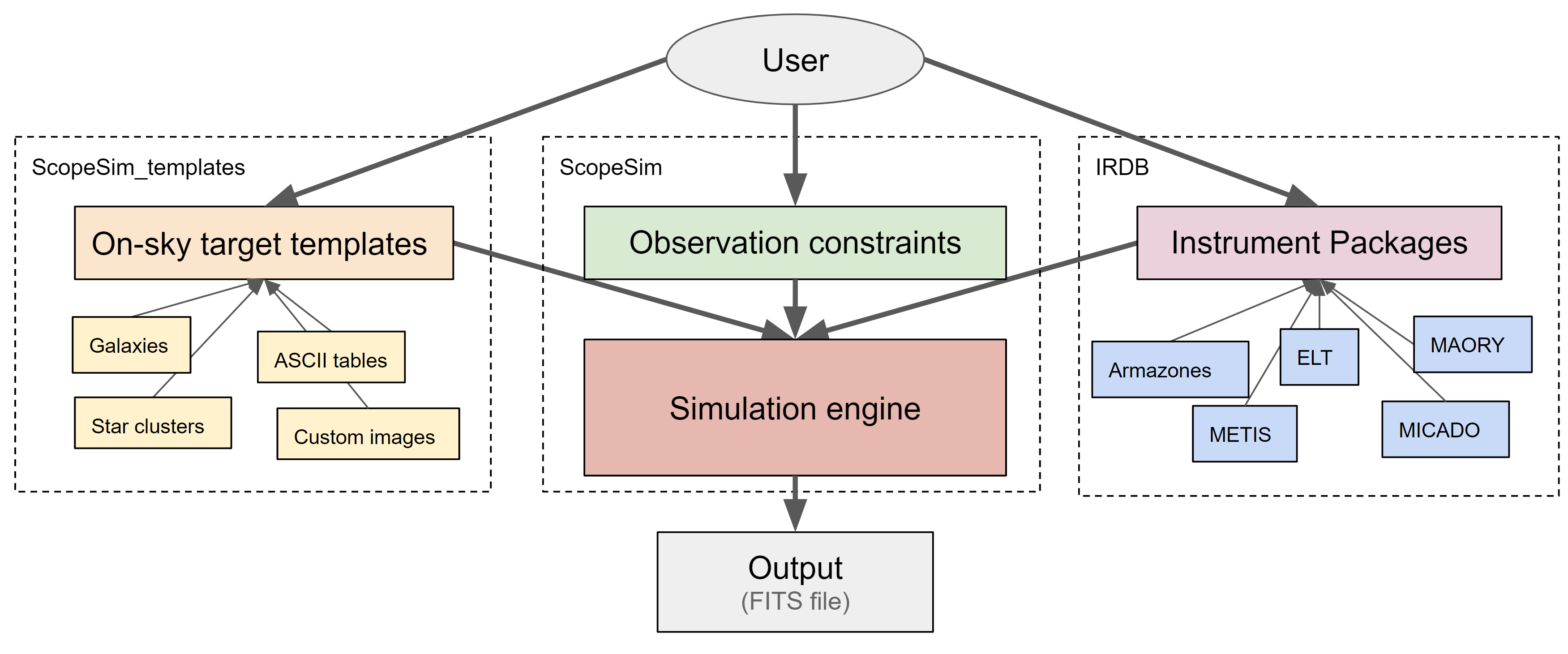}}
\caption{An illustration of the scopes of the three main packages in the ScopeSim environment.
\lstinline{ScopeSim_templates} (left) only provides the functionality to generate descriptions of on-sky objects in the format used by ScopeSim.
The instrument reference database (right) contains the data and configuration files required to generate a model of an optical system in discrete instrument packages.
The ScopeSim engine (centre) requires input from both of these packages, along with the observation constraints (e.g.~exposure time, observing mode, etc.) in order to simulate mock observation data.
The final output of a simulation is one or more FITS files containing the images or spectra of the user's  target including all expected optical aberrations associated with the optical system.}
\label{fig:framework}

\end{figure}

Figure~\ref{fig:framework} illustrates the relationship between these three packages.
Although there is a strict delineation between the scopes of each package, the interfaces between the packages allow them to interact almost seamlessly with each other.
This can be seen in the code examples from Section~\ref{examples}.
The following subsections briefly describe each of these three packages.

\subsection{ScopeSim: the observation simulator engine}
\label{scopesim-the-observation-simulator-engine}

The ScopeSim core package, also referred to as the ScopeSim engine, contains the code necessary for running observations simulations.
The code has been written in such a way as to be completely agnostic to the instrument setup as well as agnostic towards the on-sky target.
At its heart the code transports flux from a description of the on-sky target to a detector focal plane.
During the process it applies any optical aberrations contained in the optical model to the flux description.
Section~\ref{scopesim-architecture} describes this process in more detail.
ScopeSim attempts to remove as much redundancy and inefficiency from observation simulations as possible by recognising the fact that there is little need to redo the majority of the calculations executed in high-fidelity simulations.
In other words, the ScopeSim engine does not work with ray-tracing methods or use Fourier optics, except in isolated cases.
Instead it uses the fact that the observed image is a linear combination of independent optical aberrations applied to an incoming spectro-spatial flux distribution.

This focus on removing as many redundant calculations as possible results in very quick execution times.
The images from the code examples were generated on a standard laptop in around 10~seconds.
Such speed makes ScopeSim suitable for use cases with short iteration times, such a quick look feasibility studies, e.g.~``playing'' with a science case, or advanced exposure time calculators.
Such simulated observations should also help astronomers to improve their understanding of real observations and the intrinsic systematic uncertainties.
At the other end of the scale, ScopeSim is also useful for generating simulated raw data needed during the development of instrument data reduction pipelines.
As the ScopeSim engine takes its cues from the instrument packages, the fidelity of simulations is limited only by the number and accuracy of the Effects listed in these packages.

\subsection{ScopeSim Templates: Descriptions of on-sky targets}
\label{scopesim-templates-descriptions-of-on-sky-targets}

The \scopesimtemplates{} package provides a series of functions to help the user create a description of the on-sky targets they wish to observe in the format required by the \ScopeSim{} engine.
These helper function populate one or more instances of the \ScopeSim{} \Source{} class with the data needed to best describe the target.

\subsubsection{Format of a Source object}
\label{format-of-a-source-object}

In order to optimise memory usage the \Source{} objects split the spatial and spectral characteristics of a target.\cite{schmalzl2012}
These are held separately in two lists: fields and spectra.
A spatial field can be either a table of coordinated and flux scaling factors (e.g.~the positions of stars in cluster) or a 2D weight map (e.g.~an image of a galaxy).
Each entry in a field must reference one of the entries in the list of spectra, although there is no requirement for a one-to-one relationship between field entries and spectral list entries.
For example in the case of star cluster, all A0V stars can reference a single A0V spectrum.
\ScopeSim{} memory requirements are further reduced by taking advantage of this redundancy.

Galaxies and similar extended objects can also be adequately represented in this manner.
A galaxy generally contains populations of stars (new, old, high- or low-metallicity, etc.) and in most cases observations do not resolve individual stars.
Hence it can be assumed that if each component is represented by a unique flux weight map (i.e.~an image) then each stellar population can be represented by a single spectrum.
As an example, the \scopesimtemplates{} function \lstinline{galaxy.spiral_two_component} uses the B~filter image of the galaxy NGC\,1232L to represent the young population and the I~filter image for the old population.
Each image references a spectrum for a young or old population\cite{brown2014}.

The extreme limit for this type of representation would be the case where every single pixel in an image is associated with a unique spectrum.
An example might be the turbulent motions of gas in a star forming region, although it is still arguable that even here there will still be spectrally redundant regions.
In this case the spectral and spatial components can still be split, although the result will be that each spatial field entry will consist of an image with only a single pixel and referencing the associated spectrum from the list of spectra.
Such a use case would be particularly computationally expensive.
It is therefore highly recommended in such cases either to search for possible spectral redundancy before creating a \Source{} object from a spectral cube or to greatly reduce the size of the data cube to the most relevant spectral and spatial regions.

\subsubsection{Structure of ScopeSim templates package}
\label{structure-of-scopesim-templates-package}

Currently \ScopeSimtemplates{} splits the helper functions into categories based on the complexity of the Source object that is produced.
The \lstinline{basic} subpackage contains helper functions that are useful for quick look investigations, but which should not be used for in-depth feasibility studies.
For example the \lstinline{stars.cluster} function does not allow age or metallicity to be set.
In contrast the \lstinline{advanced} subpackage contains functions that can be useful for very specific science cases, but are not adapted for general use.

In addition to the general functions, it is possible to add helper functions for objects used by specific instruments.
The \lstinline{micado} subpackage for example contains functions that produce objects specific to the MICADO instrument at the ELT.
Community participation is most welcome to help expand the number of object templates in the \ScopeSimtemplates{} package.

\subsection{Instrument Reference Database: The data inside each optical model}
\label{instrument-reference-database-the-optical-model-data}

\ScopeSim{} aims to be a general-purpose instrument data simulator that can be used to simulate the output of many different optical systems.
To make this goal a reality it was mandatory that the \scopesim{} engine be completely instrument agnostic.
There is however a large amount of instrument specific data that is needed to accurately model the optical aberrations inherent in any optical system.
For \ScopeSim{} this data is stored in instrument packages in a separate instrument reference database (IRDB).
Instrument packages can be created for any self-contained section of an optical train.
The telescope, the atmosphere, the relay optics, and the instrument are generally assumed to be self-contained optical sections.
For small observatories like the University of Vienna's 1.5\,m telescope there is no benefit to splitting the optical elements into separate packages.
However, for large observatories like the VLT where multiple instruments can be attached to a single telescope, it makes sense to split the telescope optical section from the instrument optical section.
Not only does this avoid multiple versions of a single optical element (e.g.~telescope) becoming unsynchronised when one instrument package is updated and another is not, it also reduces the scope of responsibility for maintaining packages.
This means that instrument consortia would only be responsible for maintaining their own instrument package, while the telescope operator is responsible for maintaining the telescope package.
It also means a telescope or relay optics package can be updated without needing (theoretically) to inform the maintainers of all instrument packages that use a given subsystem.

\subsubsection{Instrument package format}
\label{instrument-package-format}

Each instrument package contains two main types of data:
\begin{enumerate}
\item A series of configuration files describing which optical aberrations should be modelled by \scopesim{}, and

\item The empirical data files needed for \scopesim{} to apply the aberrations to the incoming photon flux.
\end{enumerate}

The configuration files are written in the YAML\footnote{YAML is a human-readable data-serialization language. It is based on the JSON language and allows configuration files to be written in an manner that is easy to read by a human.} style.
They contain lists of \Effect{} object descriptions as well as global properties that are common to all \Effect{} objects in the subsystem.
The \Effect{} object descriptions must call an existing \Effect{} class from the \ScopeSim{} core package.
\Effect{} objects are discussed in more detail in Section~\ref{effects-objects}.
For the \Effect{}s that rely on external empirical data, these files must also be contained in the instrument package.
The empirical data files must be either ASCII tables or FITS images/tables.
Examples of empirical data files include the filter response curves or pre-computed sets of PSF kernels.

The raw instrument data currently resides in the instrument reference database on Github.
Periodically, or when explicitly needed, the data on this repository are compiled into packages and uploaded onto the \ScopeSim{} server.
It is from here that \ScopeSim{} downloads a package when asked to do so by the user (as seen in the code examples).
Packages are downloaded using \lstinline{Astropy}, and hence are saved locally in the \lstinline{Astropy} cache.
This allows the packages to be used offline.
Updated packages can be downloaded by either clearing the \lstinline{Astropy} cache, or by forcing \scopesim{} to redownload a package via the RC parameters.
An example of this is available in the online documentation.

For readers interested in creating their own instrument packages for a local telescope or instrument, the authors recommend looking inside the \lstinline{LFOA} (Leopold Figl Observatory for Astrophysics) package on the IRDB Github page.
This contains everything needed to simulate observations with the Viennese 1.5\,m telescope.

\subsection{Additional Support Packages}
\label{support-packages}

In addition to the core package, there are several stand alone packages which have been developed as part of the \ScopeSim{} framework. 
These package are not direct dependencies of \ScopeSim{}, but do help provide additional functionality to the simulation engine:

\begin{itemize}
\item \lstinline{AnisoCADO}: simulates SCAO PSFs for the ELT

\item \lstinline{SkyCalc_ipy}: queries the ESO skycalc service for atmospheric spectral curves

\item \lstinline{SpeXtra}: provides easy access to, and manipulation of, many well-known spectral libraries

\item \lstinline{Pyckles}: a light-weight wrapper for the Pickles (1998)\cite{pickles1998} and Brown (2014)\cite{brown2014} catalogues.
\end{itemize}

Table~\ref{tbl-list-of-packages} contains a list of the relevant links to both documentation and code-bases for these packages.

\begin{table}
\caption{Links to the open source documentation and code bases}
\label{tbl-list-of-packages}
\begin{tabularx}{\linewidth}{|l|X|X|}
\hline

Package
 &
Documentation
 &
Code base
 \\
\hline

ScopeSim
 &
\url{https://scopesim.readthedocs.io/}
 &
\url{https://github.io/astronomyk/scopesim}
 \\
\hline

ScopeSim\_templates
 &
\url{https://scopesim-templates.readthedocs.io/}
 &
\url{https://github.com/astronomyk/ScopeSim_templates}
 \\
\hline

IRDB
 &
\url{https://irdb.readthedocs.io/en/latest/}
 &
\url{https://github.com/astronomyk/IRDB}
 \\
\hline

AnisoCADO
 &
\url{https://anisocado.readthedocs.io/}
 &
\url{https://github.com/astronomyk/AnisoCADO}
 \\
\hline

SkyCalc\_ipy
 &
\url{https://skycalc-ipy.readthedocs.io/en/latest/}
 &
\url{https://github.com/astronomyk/SkyCalc_iPy}
 \\
\hline

SpeXtra
 &
\url{https://spextra.readthedocs.io/en/latest/}
 &
\url{https://github.com/miguelverdugo/speXtra}
 \\
\hline

Pyckles
 &
\url{https://pyckles.readthedocs.io/en/latest/}
 &
\url{https://github.com/astronomyk/Pyckles}
 \\
\hline
\end{tabularx}
\end{table}

\section{ScopeSim Architecture}
\label{scopesim-architecture}

In order to work as a multi-purpose optical\footnote{Optical refers to the wavelength ranges where telescopes act as ``photon buckets'' and detectors are in essence ``photon counters'', i.e.~from the near ultraviolet ($0.1\,\micron$) to the mid infrared ($30\,\micron$).} instrument simulator, \ScopeSim{} needs to be able to handle (at least) the two main types of instruments: imagers and spectrographs.

While every instrument is unique, all instruments, by virtue of their astronomical nature, have several key aspects in common.
All instruments:

\begin{itemize}
\item transport incoming photons through an optical system towards a detector (array),

\item use a limited number of optical components, e.g: mirrors, lenses, and gratings,

\item are only a single element in a combined optical train, which includes the atmosphere, telescope, and relay optics,

\item introduce a series of optical aberrations depending on the configuration of the optical system,

\item are generally built to behave in a predictable and repeatable manner.
\end{itemize}

These five points are important to recognise, as they have the following consequences:

\begin{itemize}
\item each optical element is responsible for one or more optical aberrations, which are not dependent on the aberrations inherent to the other optical elements,

\item the effect of each aberration on the spatial and spectral distribution of photons remains constant for a given optical configuration,

\item this constancy means the characteristics of these effects need only be calculated once and can be described by an analytical function, or an empirical data set,

\item common elements (e.g.~telescopes, atmospheres, etc.) of complex optical trains can be re-used with different instruments to create new combined optical systems.
\end{itemize}

This list of consequences implies that the final observed image from a telescope/instrument optical system is simply the sum of a discrete number of independent optical effects repositioning the incoming photons on the focal plane.

While this conclusion may seem obvious and trivial, by using it as the basis for \ScopeSim{}, it has allowed us to design and build a flexible, lightweight, general purpose instrument simulator that is capable of simulating the majority of current and future optical astronomical instruments.
\ScopeSim{} is able to mimic the optical aberrations seen in imagers, long-slit and multi-object spectrographs, as well as integral-field spectrographs.
The architecture could also theoretically be used to simulate high-contrast and high-time-resolution imagers, however these systems have not yet been tested.

\subsection{Simulation workflow}
\label{simulation-workflow}

The main \ScopeSim{} engine architecture is based around five major Python classes:

\begin{itemize}
\item \textbf{\Source{}}: holds a spectro-spatial description of the on-sky target.

\item \textbf{\FieldOfView{}}: extracts quasi-monochromatic flux maps from a \Source{} object and projects these into focal plane coordinates.

\item \textbf{\ImagePlane{}}: mimics the focal plane and acts like a 2D canvas for collecting the flux maps held in the \FieldOfView{} objects.

\item \textbf{\DetectorArray}: mimics the functionality of the instrument detector array in converting the final expectation flux image from the \ImagePlane{} into FITS format pixel maps similar to those delivered by the system's read-out electronics.

\item \textbf{\Effect{}}: the interface base class for introducing spectral, spatial, electronic, and statistical aberrations into the final flux map.
\end{itemize}

\begin{figure}

\resizebox{\linewidth}{!}{\includegraphics{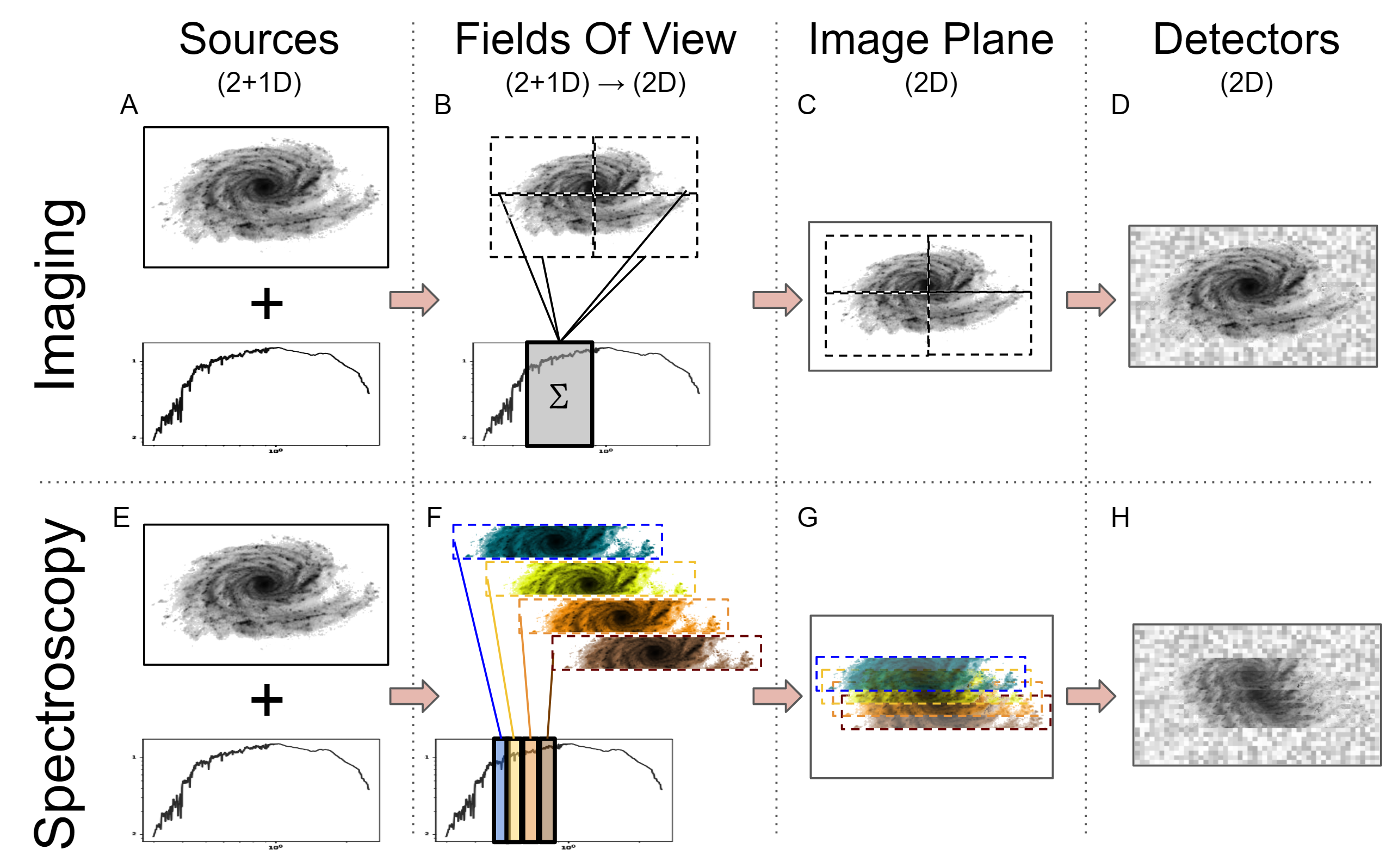}}
\caption{An illustration of the connections between the main internal classes in \ScopeSim{}: \Source{}, \FieldOfView{}, \ImagePlane{}, \DetectorArray{}.
The upper panels show the work flow for imaging simulations.
The lower panels show the work flow for spectroscopy simulations.
The work flow is in principle the same for both types of simulation.
(A, E)~Both modes require a 2+1D description of the on-sky target(s) containing linked spatial~(2D) and spectral~(1D) information.
The main difference lies in how and where the spatial and spectral borders for each \FieldOfView{} object are set.
\FieldOfView{} objects~(B, E) extract (2D)~integrated photon maps from the \Source{} object(s) and project these onto an \ImagePlane{} object.
This creates a normalised expectation image, similar to what happens at the detector focal plane in a real instrument.
The \DetectorArray{}~(D, H) extracts the regions of the \ImagePlane{} that each detector chip would see.
Simulation output in both imaging and spectroscopic cases is the same: A FITS file with detector read images in the same format as generated by the real instrument.}
\label{fig:workflow}

\end{figure}

Figure~\ref{fig:workflow} illustrates how the first four of these classes interact with each other.
The \Effect{} class is described separately in Section \ref{effects-objects}.

The \Source{} objects~(A, E) are supplied by the user.
These contain a 2+1D description of the on-sky target(s).
The spatial (2D) information is stored either as tables (collection of point sources) or as \lstinline{ImageHDU} objects (for extended objects).
Each of the spatial ``fields'' must be accompanied by one or more unique spectrum.
There need not be a one-to-one relationship between the spatial and spectral inputs.
Multiple spatial fields can reference a single spectrum.
In doing so, \ScopeSim{} can vastly reduce the amount of data that needs to be processed.
For example, a star cluster will contain many thousands of point sources.
However, only several tens of spectra are needed to adequately describe all the stars in the cluster.
There will be many hundreds of M-type stars that can reference a single common M-type stellar spectrum.

\ScopeSim{} builds a model of the optical train by importing instrument packages.
Based on the list of \Effect{}s contained in the configuration files, \ScopeSim{} splits the full spectral and spatial parameter space of the instrument into 3D ``puzzle'' pieces, known as \FieldOfView{} objects.
Each \FieldOfView{} object~(B, F) then extracts only the flux from the \Source{} object that fits within its spectro-spatial limits.
This process essentially creates a series of quasi-monochromatic puzzle pieces from the 2+1D source flux.
The spatial size and spectral depth of each puzzle piece is determined by which optical effects are included in the optical model.
For imager instruments where chromatic effects rarely play a large role, the spectral depth of each \FieldOfView{} object will be relatively large.
It is not uncommon for the spectral depth to be equal to the width of the filter bandpass.
The on-sky area for an imager is generally very large and so the viewing area is split into multiple pieces.
This is illustrated in the upper half of panel~B in Figure~\ref{fig:workflow}.
For spectrographs, the spatial component is generally small (e.g.~long slits, MOS fibre heads).
The spectral space however must be very finely sampled to accurately reproduce the spectral traces on the focal plane.
Spectrographic optical systems therefore contain many \FieldOfView{} objects which cover the same on-sky spatial region, yet cover very shallow and unique spectral ranges.
The \FieldOfView{} objects also contain two sets of spatial coordinates which connect the object's position on sky (in units of arcseconds from the optical axis) to the projected position on the detector focal plane (in units of millimeters from the optical axis).

The \ImagePlane{}~(C, G) inside each optical model acts as a 2D canvas for the integrated flux contained inside the \FieldOfView{} objects.
When each \FieldOfView{} object deposits its flux map onto the \ImagePlane{}, it simply adds the photon counts to what is already on the canvas at the \FieldOfView{}'s projected focal plane position.
The resulting \ImagePlane{} image is therefore the final integrated projected expectation flux map as would exist at the detector focal plane of a real image, in units of $\mathrm{ph\,s^{-1}\,pixel^{-1}}$.
All information on telescope aperture, viewing angle, and spectral bandpass has been integrated into the normalised photon count map.
At this stage of the simulation all sources of background flux (atmospheric, thermal) have also been projected onto the \ImagePlane{}, but no noise characteristics are included.

The \DetectorArray{} class contains a list of \Detector{} objects (D, H).
\Detector{} objects extract a region of the \ImagePlane{}'s expectation flux map corresponding to its own footprint on the detector focal plane and scales this to match the user's desired exposure time (DIT in seconds).
The resulting image is a Poisson shot noise realisation created from the expectation flux map with added detector noise characteristics, e.g.~read noise, dark current, etc.
The final detector output is returned in the form of a FITS \lstinline{HDUList} object.

\subsection{Effect Objects}
\label{effects-objects}

\begin{figure}

\resizebox{\linewidth}{!}{\includegraphics{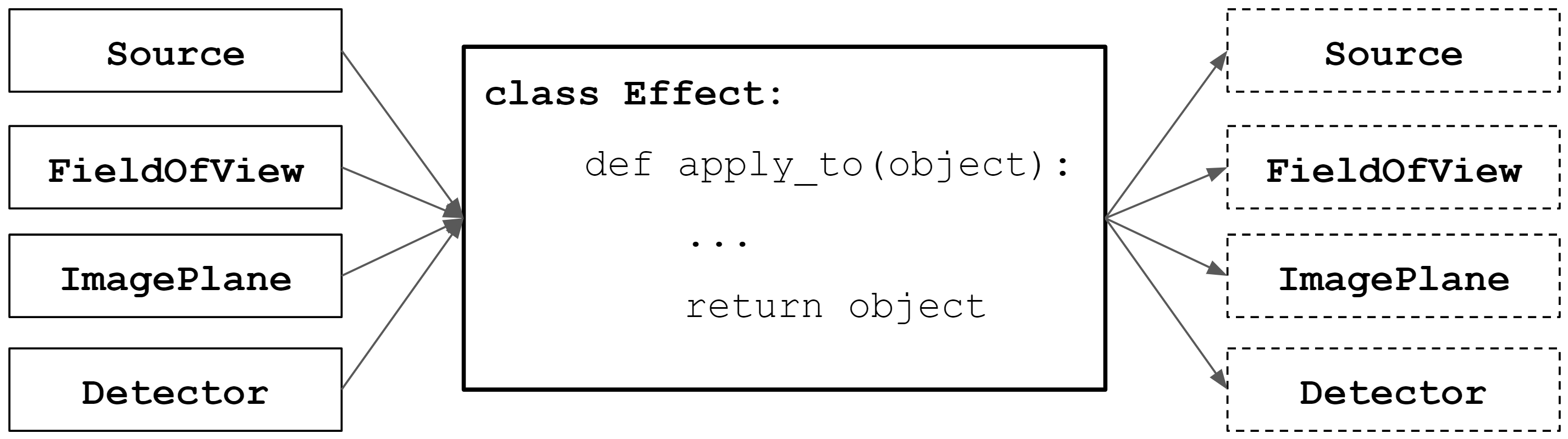}}
\caption{\Effect{} objects modify the contents of their input objects but preserve their class.
Each \Effect{} object has a single point of entry: the \lstinline{apply_to} method, which can accept any one of the four major \ScopeSim{} classes.
This method is responsible for applying optical aberrations to the flux distribution contained within those four major flux container classes.}
\label{fig:effect}

\end{figure}

A further special (and arguably the most important) \ScopeSim{} class is the \Effect{} object.
\Effect{} objects are responsible for applying any and all optical aberrations to the flux descriptions contained in the other four major flux container classes.
\Effect{} objects can contain code to alter the flux descriptions in a multitude of manners, from simple 0D alterations like adding a dark current to each pixel, to the 3D chromatic shear caused by atmospheric refraction.
In short, \Effect{} objects can be classified according to the dimensionality of their alterations to the flux descriptions:

\begin{itemize}
\item 3D: Effects are spatially and spectrally dependent aberrations, e.g: the broadband point spread function, atmospheric diffraction, etc.,

\item 2D: Effects are only spatially dependent, e.g: telescope vibration and wind shake, pupil tracking rotations, etc.,

\item 1D: Effects are only spectrally dependent, e.g: reflection and transmission curves, quantum efficiency, etc.,

\item 0D: Effects are spectrally and spatially independent. This are primarily effects that are related to photons counts and electronic noise sources, e.g:
Poisson shot noise, read-out noise, exposure stacking, detector linearity, etc.
\end{itemize}

Higher dimensional Effects are also possible albeit very rare, e.g. field varying PSFs.

Functionally, the \Effect{} class is an ``endomorphic'' operator class.
This dictates that an \Effect{} object may alter the contents of an input object, but it must not alter the input object's class type.
In other words, if a \Source{} object is the input to an \Effect{} object's \lstinline{apply_to} function, then a \Source{} object will also be returned.
This is illustrated in Figure~\ref{fig:effect}.

During the simulation workflow, the target object flux makes its way through the four main class objects described in Section~\ref{simulation-workflow}.
While flux resides in each of these objects, the relevant \Effect{}s are sequentially applied to said object.
For example, the telescope's (chromatic) PSF is applied to each of the \FieldOfView{} objects, as this is a spectrally dependent spatial (3D) effect.
In contrast, the wind-shake gaussian PSF has no spectral dependency and is therefore only applied to the \ImagePlane{}.

The following pseudo-code snippet describes the major steps of the simulation workflow and illustrates how and when the Effect objects interact with the four major flux container classes:

\begin{lstlisting}[frame=single]
source = deepcopy(orig_source)

for effect in source_effects:
    source = effect.apply_to(source)

fov.extract_from(source)

for effect in fov_effects:
    fov = effect.apply_to(fov)

image_plane.add(fov)

for effect image_plane_effects:
    image_plane = effect.apply_to(image_plane)

detector.extract_from(image_plane)

for effect detector_effects:
    detector = effect.apply_to(detector)

detector.write_to("file.fits")
\end{lstlisting}

As can be seen, there is a very similar pattern.
Obviously there are a few more steps involved in the actual \ScopeSim{} code, however the \lstinline{observe} method of an optical model consists of little more than a Python implementation of this pseudo-code.

\ScopeSim{} already includes a large number of standard \Effect{}s in the core package, including, but not limited to: PSFs, transmission curves, Filter wheels, atmospheric dispersion, detector readout noise models, etc (see online documentation for a full list).
It is clear however that there are many more that could be added.
The \Effect{} object interface has been intentionally kept light weight to encourage users to implement custom effects for their own simulations.
The online documentation contains a tutorial on how to write custom effects.
Users are also cordially invited to submit any custom \Effect{}s they deem useful to the wider community to the \ScopeSim{} package as a pull request via the Github repository.

\section{Conclusion}
\label{conclusion}

\ScopeSim{} is a flexible multipurpose instrument data simulation framework built in Python.
It enables both raw and ideal observation data to be simulated for a wide range of telescopes and instruments quickly and efficiently on standard personal computers.
This is achieved by keeping the instrument model data, descriptions of the target objects, and simulation engine strictly separated.
The three main packages are the \ScopeSim{} engine, a library of target templates, and the instrument reference database.
Several existing and future telescope and instrument systems have been already been implemented, with more to come in the future.
For example, work is steadily progressing on the instrument packages for the MICADO and METIS\cite{metis2018} instruments at the ELT.

\textbf{Community involvement is highly encouraged!}
The whole ScopeSim framework is open source and the developers welcome any contributions, both code and comments, by members of the astronomical community.
The astronomical object templates package is one area which will benefit greatly from community contributions.
There is a wide variety of astronomical objects for which the authors have not yet created templates.
Galaxy clusters, gravitational lenses, supernovae, exoplanets, solar system objects are all still missing from the \ScopeSimtemplates{} package.
The instrument reference database also currently only contains the instruments directly relevant to the authors, i.e. MICADO, METIS, HAWKI, and the LFOA.
There is no limit to the size of telescopes or number of instruments that can be hosted on the server.
Readers interested in submitting a package for their own telescope or instrument are very welcome to make a pull request on the \IRDB{} Github page.

\acknowledgments 
\ScopeSim{} incorporates parts of Bernhard Rauscher's HxRG Noise Generator package for Python\cite{nghxrg}.
\ScopeSim{} uses the following common 3rd party python packages as dependecies: Numpy\cite{numpy}, Scipy\cite{scipy}, Astropy\cite{astropy1, astropy2}, Synphot\cite{synphot}, Pyyaml, AnisoCADO\cite{anisocado}.
Optional dependencies include: Matplotlib\cite{matplotlib}, Jupyter\cite{jupyter}
\ScopeSim{} makes use of atmospheric transmission and emission curves generated by ESO's SkyCalc service, which was developed at the University of Innsbruck as part of an Austrian in-kind contribution to ESO\cite{skycalc1, skycalc2}.
This research is partially funded by the project IS538004 of the Hochschulraumstrukturmittel (HRSM) provided by the Austrian Government and administered by the University of Vienna.
The authors would also like to thank all the members of the consortium for their effort in the MICADO project, and their contributions to the development of this tool.

\bibliography{main} 

\end{document}